\documentclass[fp,twocolumn]{jpsj3}
\usepackage{graphicx}
\usepackage{xcolor}
\usepackage{latexsym}
\usepackage{amsmath,amssymb}
\usepackage{braket}
\usepackage{cite}
\usepackage{url}

\title{Evaluating the solution performance of the augmented Lagrangian function on Ising machines}

\author{Shunsuke Awai$^{1}$\thanks{shunsuke.awai@keio.jp}, Takuro Itoh$^{1}$, Keita Takahashi$^{1}$, Kotaro Tanahashi$^{2}$, Shu Tanaka$^{1,3,4,5}$\thanks{shu.tanaka@keio.jp}}

\inst{
$^1$Graduate School of Science and Technology, Keio University, 3-14-1 Hiyoshi, Kohoku-ku, Yokohama-shi, Kanagawa 223-8522, Japan\\
$^2$Recruit Co., Ltd., 1-9-2 Marunouchi, Chiyoda-ku, Tokyo 100-6640, Japan\\
$^3$Department of Applied Physics and Physico-Informatics, Keio University, 3-14-1 Hiyoshi, Kohoku-ku, Yokohama-shi, Kanagawa 223-8522, Japan\\
$^4$Keio University Sustainable Quantum Artificial Intelligence Center (KSQAIC), Keio University, 2-15-45, Mita, Minato-ku, Tokyo 108-8345, Japan\\
$^5$Human Biology-Microbiome-Quantum Research Center (WPI-Bio2Q), Keio University, 35 Shinanomachi, Shinjuku-ku, Tokyo 160-8582, Japan\\} 

\abst{
We apply the augmented Lagrangian function (ALF) as a formulation for Ising machines and evaluate its performance by time-to-epsilon ($\mathrm{TT\varepsilon}$). The ALF has been well studied in continuous optimization for its numerical stability and convergence, and its advantage over the penalty function formulation is demonstrated here through the following results. Using the quadratic knapsack problem as a benchmark, we examine the dependence of $\mathrm{TT\varepsilon}$ on the hyperparameters $\mu$ and $\lambda$. The augmented Lagrangian formulation reduces $\mathrm{TT\varepsilon}$ by roughly an order of magnitude compared with the penalty function formulation, keeping $\mu$ small while obtaining feasible solutions and, for representative parameter settings, reaching high-precision solutions earlier in the search. These findings indicate that the augmented Lagrangian function is a promising formulation for improving the solution performance of Ising machines.
}

\begin{document}

\maketitle

\section{Introduction}
\label{Sec:Introduction}

Combinatorial optimization is a mathematical problem of finding a combination of decision variables that minimizes or maximizes an objective function under given constraints~\cite{karp2009reducibility}. 
Many real-world problems, such as logistics optimization, manufacturing scheduling, and communication routing, can be reduced to this form~\cite{dantzig1959truck, laporte2009fifty, magnanti1984network, alonso2017demand}.
As the number of decision variables increases, the search space grows exponentially, and exhaustive search becomes intractable for large-scale problems.
Developing efficient solution methods for such problems is therefore a long-standing and important challenge.

To address this challenge, Ising machines have attracted attention as computing devices capable of efficiently searching for good solutions to combinatorial optimization problems\cite{mohseni2022ising}, and they have been applied to various practical problems such as financial optimization~\cite{rosenberg2015solving, Tatsumura2023, takahashi2025effectiveness}, advertising optimization~\cite{Tanahashi2019}, logistics~\cite{mukasa2021ising, weinberg2023supply, kanai2024annealing, kawase2026parallelizable}, materials science~\cite{Harris2018SpinGlass, King2018Topological, kitai2020designing, Utimula2021Ionic, sampei2023quantum, couzinie2025machine}, computer-aided engineering~\cite{endo2022phase, honda2024development, kondo2025simultaneous}, and biology~\cite{perdomo2012finding, irback2022folding, kikuchi2026highorder, kikuchi2026factorization}.
In an Ising machine, the target combinatorial optimization problem is formulated as an Ising model, a mathematical model in statistical mechanics. 
The ground state of this model is then explored using state transitions based on physical or algorithmic dynamics. 
Representative internal algorithms of Ising machines include those based on simulated annealing (SA) and quantum annealing (QA). 
In SA-based Ising machines, a temperature parameter is introduced as a thermal fluctuation effect and gradually reduced to search for the ground state of the Ising model~\cite{kirkpatrick1983optimization,johnson1989optimization,johnson1991optimization}.
In QA-based Ising machines, quantum fluctuations are used to induce tunneling between states, and the ground state is searched for by gradually reducing the strength of these fluctuations~\cite{kadowaki1998quantum}.
Because Ising machines exhibit such stochastic behavior, the solution obtained can differ from trial to trial even when the same problem is solved under identical conditions.

When solving a constrained combinatorial optimization problem with an Ising machine, a constraint term representing the constraints is added to the objective function, thereby casting the problem into an unconstrained form. 
Here, the penalty coefficient $\mu$, which balances the relative contributions of the objective function and the constraints, must be set to an appropriate value. 
The penalty coefficient $\mu$ involves an inherent trade-off. 
If $\mu$ is too small, infeasible states can have low energies, and the probability of sampling feasible solutions decreases.  
If $\mu$ is too large, the energy landscape is dominated by the constraint term. 
In finite-time heuristic searches on Ising machines, this dominance can weaken the effect of objective-value differences, and feasible solutions with small objective error become difficult to obtain~\cite{takehara2019multiple,ayodele2022penalty}.
The setting of $\mu$ that balances constraint satisfaction and solution precision is thus a key challenge.

To address this trade-off, numerous tuning methods for finding an appropriate $\mu$
have been reported~\cite{rosenberg2015solving, takehara2019multiple, ayodele2022penalty, yin2024penalty, ide2025extending, qin2026variational, ide2026unfair}.
One representative approach increases $\mu$ incrementally until a feasible solution is obtained~\cite{rosenberg2015solving}.
Another sets $\mu$ based on an upper bound of the objective function value~\cite{ayodele2022penalty}, but the resulting $\mu$ tends to be excessively large.
As another approach, for the traveling salesman problem, the search range is defined by the minimum and maximum distances between any two points. 
Within this range, an approach that tries multiple penalty coefficients has also been proposed~\cite{takehara2019multiple}.
However, all of these methods share a common limitation: they adjust the trade-off using a single parameter.

A natural formulation for mitigating this trade-off is the augmented Lagrangian function (ALF). 
The ALF introduces a Lagrange multiplier term in addition to the quadratic penalty term. 
In continuous optimization, augmented Lagrangian methods are known to alleviate the ill-conditioning associated with excessively large penalty coefficients~\cite{hestenes1969multiplier}. 
In the present Ising-machine setting, we do not rely on this continuous-optimization convergence theory. 
Instead, we use the ALF as a QUBO formulation and examine whether the Lagrange multiplier term enables feasible solutions to be sampled while keeping $\mu$ relatively small. 
In this sense, the ALF is expected to preserve the contribution of the objective term more effectively than the penalty function.

Previous studies on the augmented Lagrangian approach for Ising machines have addressed two aspects: formulation efficiency and algorithmic efficiency.
On the formulation side, the augmented Lagrangian formulation reduced the number of auxiliary variables required for inequality constraints, enabling the solution of larger-scale problems~\cite{djidjev2023quantum, cellini2024qal, hong2025qubit}.
On the algorithmic side, the augmented Lagrangian method (ALM), which adaptively updates the hyperparameters based on the constraint violation, reduced the number of parameter updates compared with the conventional penalty method~\cite{tanahashi2021augmented, djidjev2023logical}.
However, these studies did not directly evaluate how the ALF as a formulation influences the solution performance of the Ising machine.
The objective of this study is therefore to evaluate the ALF as a formulation rather than as an iterative parameter-update algorithm.

Using the quadratic knapsack problem as a benchmark, we formulate it with both the penalty function and the ALF on the same QUBO basis and compare their performance under two analyses.
First, we quantify the dependence of $\mathrm{TT\varepsilon}$ on $\mu$ and $\lambda$ and identify the parameter region in which the ALF achieves a lower $\mathrm{TT\varepsilon}$ than the penalty function. 
Second, we examine the temporal evolution of the incumbent solution within the execution time to clarify how the formulation affects the early stage of the search. 
For the benchmark instances tested in this study, the ALF reduces $\mathrm{TT\varepsilon}$ by roughly an order of magnitude compared with the penalty function.
Furthermore, through a square-completion analysis, we show that the Lagrange multiplier term shifts the effective capacity of the constraint, providing a structural explanation for the observed improvement.

The remainder of this paper is organized as follows.
Section~\ref{sec:constraint} introduces the input format for the Ising machine used in this study. 
Section~\ref{sec:method} defines the formulation methods adopted and the evaluation metric. 
Section~\ref{sec:setup} describes the specific experimental conditions in detail. 
Section~\ref{sec:results} presents the results obtained. 
Section~\ref{sec:discussion} discusses these results, and Sect.~\ref{sec:conclusion} concludes this study.
\section{Input Format for the Ising Machine}
\label{sec:constraint}

This section presents the form of the energy function that is input to the Ising machine, in order to clarify the formulation of the constraint-handling methods described later. 
First, Subsect.~\ref{subsec:ising_model} describes the Ising model, which is the input format of the Ising machine. 
Then, Subsect.~\ref{subsec:QUBO} introduces Quadratic Unconstrained Binary Optimization (QUBO), which is known as a representation mathematically equivalent to the Ising model.

\subsection{Ising Model}
\label{subsec:ising_model}

The Ising model is a mathematical model in statistical mechanics that represents cooperative phenomena in strongly correlated systems, and it is defined on an undirected graph $G = (V, E)$~\cite{ising1925beitrag}.
Here, $V$ denotes the set of vertices, and $E$ denotes the set of undirected edges representing the connections between vertices. 
A variable $s_i \in \{+1, -1\}$ that takes one of two states, up ($+1$) or down ($-1$), is defined on each vertex $i \in V$, and it is called a spin. 
An interaction $J_{i,j}$ exists between adjacent spins $s_i$ and $s_j$. 
In addition, a local external magnetic field $h_i$ is assumed to be applied to each spin $s_i$. 
The case $J_{i,j} > 0$ is called a ferromagnetic interaction, and the case $J_{i,j} < 0$ is called an antiferromagnetic interaction. 
Here, $J_{i,j}$ and $h_i$ are real constants. The Hamiltonian $H_{\mathrm{Ising}}$ of this system is expressed as follows.
\begin{equation}
    H_{\mathrm{Ising}} = -\sum_{(i,j) \in E} J_{i,j} s_i s_j - \sum_{i \in V} h_i s_i, \quad s_i \in \{+1, -1\}.
    \label{eq:Isingmodel}
\end{equation}
The first term on the right-hand side of Eq.~\eqref{eq:Isingmodel} represents the interaction energy, and the second term represents the potential energy due to the local magnetic fields. 
When a combinatorial optimization problem is formulated as an Ising model, the Hamiltonian is constructed so that lower-energy spin configurations correspond to better solutions.
If the formulation is exact, the ground state corresponds to an optimal solution.
An Ising machine attempts to solve the combinatorial optimization problem by searching for low-energy configurations, ideally the ground state, of this Hamiltonian.

\subsection{Quadratic Unconstrained Binary Optimization (QUBO)}
\label{subsec:QUBO}

Quadratic Unconstrained Binary Optimization (QUBO) is defined on the same undirected graph $G=(V,E)$ using binary variables $x_i$ that take the value 0 or 1.
The binary variable $x_i$ is defined on vertex $i$.
The weight of the edge between vertices $i$ and $j$ is denoted by $a_{i,j}$, and the weight of vertex $i$ is denoted by $b_{i}$.
Note that $a_{i, j}$ and $b_{i}$ are real constants.
The Hamiltonian $H_{\mathrm{QUBO}}$ of the QUBO is then expressed as follows.
\begin{equation}
    H_{\mathrm{QUBO}} = - \sum_{(i,j) \in E} a_{i,j} x_i x_j-\sum_{i\in V} b_i x_i,\quad x_i \in \{0, 1\}.
    \label{QUBO}
\end{equation}
Diagonal quadratic terms are absorbed into the linear coefficients because $x_i^2 = x_i$ for binary variables.
The variable $s_i$ of the Ising model takes the value $\pm1$, and the variable $x_i$ of the QUBO takes the value 0 or 1.
From the following relation, the two variables are mutually convertible, and accordingly the Ising model and the QUBO are mathematically equivalent.
\begin{equation}
    x_i = \frac{s_i+1}{2}.
    \label{IsingequaltoQUBO}
\end{equation}
Substituting Eq.~\eqref{IsingequaltoQUBO} into Eq.~\eqref{QUBO} recovers the Ising form up to an additive constant. This constant shifts the absolute energy but does not change the ground-state configuration.
In this study, the objective and constraint terms introduced in the following sections are expressed in the QUBO representation. 
This representation serves as the common basis for comparing the penalty function and the augmented Lagrangian function.

\section{Method}
\label{sec:method}
The main comparison in this study is between the penalty function and the augmented Lagrangian function at fixed hyperparameters. The penalty method and the augmented Lagrangian method are described to clarify their relation to iterative hyperparameter tuning, and the augmented Lagrangian method is examined as an auxiliary analysis in Sect.~\ref{sec:discussion}.

This section describes the method for evaluating how the formulation of the constraints influences the solution performance of the Ising machine.
The QUBO representation introduced in the previous section is used as a common input representation, and only the way the constraints are formulated is changed.
The structure of this section is as follows.
Subsection~\ref{subsec:formulation} describes the penalty function, which converts a constrained combinatorial optimization problem into an unconstrained form, and the augmented Lagrangian function adopted in this study.
Then, Subsect.~\ref{subsec:TTepsilon} introduces $\mathrm{TT\varepsilon}$, which is adopted as the evaluation metric in this study.

\subsection{Compared Formulation Methods}
\label{subsec:formulation}
This subsection describes the formulation methods used in this study.
In all methods, the QUBO representation and the problem setting are kept common, and only the handling of the constraints is changed.
This makes it possible to compare the influence of the difference in formulation on the solution performance.
First, the commonly used formulation called the penalty function is described.
Then, the augmented Lagrangian function, which is formulated by adding a term linear in the constraint to the penalty function, is described.

\subsubsection{Penalty Function}
\label{subsubsec:penalty_function}
The penalty function, which is a QUBO formulation method for constrained combinatorial optimization problems, is described.
First, the state vector whose elements are the $n$ binary variables $x_i \in \{0,1\}$ is defined as $\boldsymbol{x} = (x_1, x_2, \dots, x_n)^{\mathrm{T}}$.
Let the objective function be $f(\boldsymbol{x})$ and the imposed equality constraint be $g(\boldsymbol{x}) = c$, where $c$ is a real constant.
The problem Hamiltonian $H_{\mathrm{P}}(\boldsymbol{x})$ using the penalty function is then given by the following equation.
\begin{equation}
    H_{\mathrm{P}}(\boldsymbol{x}) = f(\boldsymbol{x}) + \frac{\mu}{2} \left[g(\boldsymbol{x}) - c\right]^2.
    \label{eq:penalty_example}
\end{equation}
The first term on the right-hand side is the objective function to be minimized, and the second term is the penalty term that enforces the constraint.
Here, $H_{\mathrm{P}}(\boldsymbol{x})$ is a QUBO when the objective is at most quadratic and the equality constraint $g(\boldsymbol{x}) = c$ is affine in the binary variables. 
This condition holds for the quadratic knapsack problem treated in this study.
The penalty coefficient $\mu\,(>0)$ is a hyperparameter that determines the energy scale of the constraint term relative to the objective term.
If the penalty coefficient is too small, infeasible states can have low energies, and the probability of sampling feasible solutions decreases. 
By contrast, if the penalty coefficient is too large, the energy landscape is dominated by the constraint term. 
In finite-time heuristic searches on Ising machines, this dominance can weaken the effect of objective-value differences, and feasible solutions with small objective error become difficult to obtain~\cite{ayodele2022penalty}.

\subsubsection{Penalty Method}
\label{subsubsec:penalty_method}
The appropriate value of the penalty coefficient $\mu$ differs from problem to problem and is difficult to specify in advance.
Therefore, a method that does not fix $\mu$ but increases it incrementally through iterative computation is commonly used~\cite{rosenberg2015solving}.
This is called the penalty method.
In the penalty method, the coefficient is increased according to the following update rule, starting from a small initial value $\mu^{(0)} > 0$.
\begin{equation}
    \mu^{(k+1)} \leftarrow \alpha \mu^{(k)} \quad (\alpha >1, k=0, 1, 2, ...),
    \label{eq:mu_update}
\end{equation}
where $k$ is the iteration index, and $\alpha$ is the growth factor of the penalty coefficient $\mu$.
This algorithm terminates when a feasible solution is obtained, or when the iteration index $k$ reaches a predetermined upper limit.
The settings of the growth factor $\alpha$ and the initial value $\mu^{(0)}$ affect the solution performance.
If these are too large, $\mu$ increases too quickly.
As a result, the energy scale of the objective function term becomes small relative to the constraint term, which increases the objective error of the obtained solution.
By contrast, if these are too small, many iterations are required to obtain a feasible solution.

\subsubsection{Augmented Lagrangian Function}
\label{subsubsec:augmented_Lagrangian_function}

The augmented Lagrangian function, which combines the concept of the method of Lagrange multipliers with the penalty function introduced in Subsubsect. \ref{subsubsec:penalty_function}, is described.
For the objective function $f(\boldsymbol{x})$ and the equality constraint $g(\boldsymbol{x}) = c$ as in Subsubsect. \ref{subsubsec:penalty_function}, the problem Hamiltonian $H_{\mathrm{AL}}(\boldsymbol{x})$ using the augmented Lagrangian function is given by the following equation.
\begin{equation}
    H_{\mathrm{AL}}(\boldsymbol{x}) = f(\boldsymbol{x}) + \frac{\mu}{2} \left[g(\boldsymbol{x}) - c\right]^2 - \lambda \left[ g(\boldsymbol{x}) - c\right].
    \label{eq:AL_example}
\end{equation}

A key feature of the augmented Lagrangian function is that, in addition to the quadratic term based on the penalty function (the second term on the right-hand side), it introduces a linear term (the third term on the right-hand side) that uses the Lagrange multiplier $\lambda$ as a new hyperparameter.
The linear term plays the role of correcting the energy according to the amount of constraint violation.
For this correction to work as intended, $\lambda$ must be chosen appropriately.
We formulate the following hypothesis.
The introduction of the Lagrange multiplier term $\lambda$ makes it possible to keep the penalty coefficient $\mu$ required to obtain a feasible solution relatively smaller than in the formulation using the penalty function.
As a result, the energy scale of the objective function term relative to the constraint term is maintained.
Consequently, the probability of obtaining feasible solutions with small objective error is expected to improve.

\subsubsection{Augmented Lagrangian Method}
As in the penalty method, iterative parameter updates can also be performed in the formulation using the augmented Lagrangian function.
This is called the augmented Lagrangian method.
The major difference from the penalty method is that, at each iteration step, the Lagrange multiplier $\lambda$ is updated according to the following.
\begin{align}
    \label{eq:lambda_update}
    &\lambda^{\left( k+1 \right)} \leftarrow \lambda^{\left( k \right)} - \mu^{\left( k \right)}\big\langle g (\boldsymbol{x}^{\left( k \right)}) - c\big\rangle,\\
    \label{eq:mu_update_alm}
    &\mu^{\left( k+1 \right)} \leftarrow \alpha \mu^{\left( k \right)} \quad (\alpha >1, k=0, 1, 2, ...),
\end{align}
where $\big\langle g \left(\boldsymbol{x}^{\left( k \right)}\right) - c \big\rangle$ denotes the average constraint violation value of the solutions obtained at each iteration.
The Lagrange multiplier $\lambda$ is updated in a weighted manner according to this amount of constraint violation.
The larger the constraint violation, the more $\lambda$ is updated.
This sets the hyperparameter in the direction of satisfying the constraint.
Owing to this update rule, the augmented Lagrangian method has been reported to reach a feasible solution without increasing $\mu$ as excessively as the penalty method.

\subsection{Evaluation by $\mathrm{TT\varepsilon}$}
\label{subsec:TTepsilon}
In this study, the solution performance refers to the ability to obtain, within a given execution time, a solution that is feasible for the original constrained problem and whose objective error is at most $\varepsilon$.
It is quantified by time-to-epsilon ($\mathrm{TT\varepsilon}$), the total computation time required to obtain such a solution with a target probability.
Several metrics exist for evaluating the performance of computers that exhibit stochastic behavior, and they differ in whether the optimal solution is required for the evaluation.
The optimal-solution-finding probability and the time-to-solution ($\mathrm{TTS}$) are metrics that measure the performance based on the optimal solution~\cite{ronnow2014defining}.
These are useful for problems whose optimal solution is known.
However, they are applicable only when the computer can find the optimal solution with a nonzero probability. 
Therefore, they are not applicable when the problem size is so large that the optimal solution cannot be found at all. By contrast, the objective function value and the feasible-solution-finding probability can evaluate the quality of approximate solutions without requiring the optimal solution.

In this study, $\mathrm{TT\varepsilon}$ is adopted as the evaluation metric.
$\mathrm{TT\varepsilon}$ requires the optimal objective value as a reference but does not require the solver to reach the exact optimal solution, because it admits solutions within a relative error $\varepsilon$. 
It thus occupies an intermediate position between the two classes of metrics described above. 
This metric is adopted in order to evaluate, on an execution-time basis, how the difference in formulation affects the time required to reach feasible solutions with small objective error. The temporal behavior within a single run is examined separately through the incumbent-solution trace in Sect.~\ref{sec:results}. 
$\mathrm{TT\varepsilon}$ is a generalization of the time-to-solution ($\mathrm{TTS}$) metric~\cite{ronnow2014defining} to approximate optimization, in which success is defined as obtaining a solution within a prescribed relative error $\varepsilon$ of the optimum~\cite{munozbauza2025scaling}.
In this study, we extend this criterion to constrained problems by requiring that the returned solution is both feasible for the original constrained problem and within a relative error $\varepsilon$ of the optimal objective value.
For computers with stochastic behavior, the performance must be evaluated not only on the basis of a single computation time, but also on the basis of the total computation time, which integrates the execution time per run (the annealing time) and the number of trials.
In addition, for large-scale combinatorial optimization problems, finding the optimal solution can be difficult.
In such cases, a practically important metric is how quickly a solution satisfying the target relative error $\varepsilon$ can be reached.

The annealing time of a single run is denoted by $\tau$.
Here, $p_{\varepsilon}(\tau)$ denotes the probability that the solution $\boldsymbol{x}$ obtained in a single run is feasible for the original constrained problem and that its objective function value $f(\boldsymbol{x})$ satisfies a relative error of $\varepsilon$ or less with respect to the optimal objective value $f^\ast$, that is,
\begin{align}
\left| \frac{f(\boldsymbol{x}) - f^\ast}{f^\ast} \right| \le \varepsilon.
\end{align}
The probability $p_{R}$ of obtaining a solution with a relative error of $\varepsilon$ or less at least once over $R$ repeated annealing runs is expressed by the following equation.
\begin{equation}
    p_{R} = 1 - (1 - p_{\varepsilon}(\tau))^R.
    \label{eq:TTepsilon_p_r}
\end{equation}
Solving this equation for $R$ gives
\begin{equation}
     R= \frac{\ln(1 - p_{R})}{\ln(1 - p_{\varepsilon}(\tau))}.
\end{equation}
Multiplying $R$ by the computation time per run $\tau$ gives the total computation time $\mathrm{TT\varepsilon}$ required to obtain a solution satisfying a relative error of $\varepsilon$ with a probability $p_{R}$, as follows.
\begin{equation}
    \mathrm{TT\varepsilon}(\tau, p_{R}) = \tau \frac{\ln(1 - p_{R})}{\ln(1 - p_{\varepsilon}(\tau))}.
\end{equation}
Here, $p_{\varepsilon}(\tau)$ is estimated from the empirical success probability over the repeated trials. When $p_{\varepsilon}(\tau)=1$, a single run suffices and $\mathrm{TT\varepsilon}=\tau$. 
When $p_{\varepsilon}(\tau)=0$, no successful solution is observed and $\mathrm{TT\varepsilon}$ is undefined. 

The Ising model and the QUBO are equivalent through a variable transformation, but the constant term changes with the transformation, so the value of $f^*$ differs between the two representations. 
Because $\mathrm{TT\varepsilon}$ is defined through the relative error with respect to $f^*$, its value depends on the representation used.
All computations in this study are unified in the QUBO form, and the consistency of the comparison is thereby maintained.
\section{Experimental Setup}
\label{sec:setup}
To test the hypothesis described in Subsubsect.~\ref{subsubsec:augmented_Lagrangian_function}, we conducted two investigations.
First, we evaluated $\mathrm{TT\varepsilon}$ over the $\lambda-\mu$ parameter space, visualized as a heatmap, to identify the parameter region in which $\mathrm{TT\varepsilon}$ is small.
Second, for representative parameter settings selected from this region, we examined the temporal evolution of the incumbent solution during a single run.
The former quantifies the parameter dependence of the formulation, whereas the latter examines how the formulation affects the early stage of the search.
Subsection~\ref{subsec:quadratic_knapsack_problem} describes the formulation of the quadratic knapsack problem, which is the inequality-constrained combinatorial optimization problem treated in this study.
Then, Subsect.~\ref{subsec:parameter_setting} describes the specific experimental conditions and hyperparameter settings used in this study.

\subsection{Problem Setting: Quadratic Knapsack Problem}
\label{subsec:quadratic_knapsack_problem}
This study targets the quadratic knapsack problem (QKP), which is a representative inequality-constrained combinatorial optimization problem.
In addition, benchmark instances whose optimal solutions are known are available.
For these reasons, the QKP is adopted as the target for verifying the effectiveness of the augmented Lagrangian function.
The objective of this problem is to select a combination of decision variables under the constraint that the capacity of the knapsack is not exceeded.
The objective is to maximize the sum of the value of each item and the mutual value between items.
The number of items is denoted by $N$, and the mutual value between items $i$ and $j$ is denoted by $p_{i,j}$.
Here, $p_{i,j}$ is an element of the interaction matrix $P$.
The diagonal elements of the interaction matrix $P$ correspond to the value of the items themselves, and the off-diagonal elements correspond to the mutual value between items.
The matrix $P$ is defined as an upper triangular matrix.
\begin{equation}
    P=
    \begin{bmatrix}
    p_{1,1} & \dots & p_{1,N}\\
    \vdots & \ddots & \vdots\\
    0&\dots&p_{N,N}
    \end{bmatrix}.
    \label{eq:density_matrix}
\end{equation}

Since the original problem maximizes the total value, its objective function is expressed as the following minimization problem by negating the total value.
\begin{equation}
    {H}_{\mathrm{obj}} = - \sum_{1\leq i \leq j \leq N}  p_{i,j} x_i x_j, \quad  x_i\in \{0, 1\},
    \label{eq:knapsack_obj}
\end{equation}
where $x_i$ is a binary variable that takes the value $1$ when item $i$ is included in the knapsack and $0$ when it is not.
As the constraint, an upper limit on the total weight of the items placed in the knapsack is set.
Letting the knapsack capacity be $W$, the constraint is expressed as the following inequality.
\begin{equation}
    \sum_{i=1}^N w_i x_i \leq W,
    \label{eq:inequality_const}
\end{equation}
where $w_i$ is the weight of item $i$.
The inequality constraint shown in Eq.~\eqref{eq:inequality_const} cannot be input directly into the Ising machine.
To address this issue, a slack variable, which is a non-negative auxiliary variable, is introduced to convert the inequality constraint into an equality constraint~\cite{lucas2014ising, tanaka2017quantum, Tanahashi2019, zaman2021pyqubo}.
Specifically, an auxiliary variable is introduced to represent the difference between the total weight and the upper limit $W$.

An integer-to-binary conversion using logarithmic scaling (log encoding) is applied.
By representing the integer value using multiple binary variables $y_d\in \{0, 1\}$, the constraint term $H_\mathrm{const}$ is defined as in Eq.~\eqref{eq:knapsack_const} below.
\begin{equation}
    {H}_{\mathrm{const}} = \sum_{d=0}^{D-1} 2^d y_d - (2^D - 1 -W) - \sum_{i=1}^{N} w_i x_i, \quad y_d\in \{0, 1\},
    \label{eq:knapsack_const}
\end{equation}
where $D=\lceil \log_2 (W+1) \rceil$, and $y_d$ denotes an auxiliary variable.
Here, $\lceil \cdot \rceil$ is the ceiling function.
The range of the binary expansion $\sum_{d=0}^{D-1} 2^d y_d$ by log encoding is $[0, 2^D - 1]$.
The binary expansion $\sum_{d=0}^{D-1} 2^d y_d$ itself does not directly represent the remaining capacity. 
With the definition in Eq.~\eqref{eq:knapsack_const}, the quantity $2^D-1-\sum_{d=0}^{D-1} 2^d y_d$ represents the encoded remaining capacity. 
For a feasible item selection satisfying Eq.~\eqref{eq:inequality_const}, there exists an assignment of the auxiliary variables such that $H_{\mathrm{const}}=0$.
For an infeasible item selection with $\sum_{i=1}^{N} w_i x_i > W, H_{\mathrm{const}}<0$ for any assignment of the auxiliary variables.

The problem Hamiltonian formulated using the penalty function described in Subsubsect.~\ref{subsubsec:penalty_function} is as follows.
\begin{equation}
    {H}_{\mathrm{P}}   =  {H}_{\mathrm{obj}} + \frac{\mu}{2}({H}_{\mathrm{const}})^2.
    \label{eq:knapsack_PM}
\end{equation}
In contrast, the problem Hamiltonian formulated using the augmented Lagrangian function described in Subsubsect.~\ref{subsubsec:augmented_Lagrangian_function} is expressed as
\begin{equation}
   {H}_{\mathrm{AL}} =  {H}_{\mathrm{obj}} + \frac{\mu}{2}\left({H}_{\mathrm{const}}\right)^2 - \lambda {H}_{\mathrm{const}}.
   \label{eq:knapsack_ALM}
\end{equation}
Because $H_{\mathrm{const}}$ is affine in the binary variables $x_i$ and $y_d$, both Hamiltonians remain quadratic functions of the binary variables.

\subsection{Experimental Conditions and Parameter Settings}
\label{subsec:parameter_setting}
\begin{table}[t]
    \centering
    \caption{Computational conditions for the parameter search}
    \begin{tabular}{l c}
        \hline
        Trials per parameter setting & $50$ \\
        Execution time [s]  & $1$ \\
        Range of $\mu$ & $0$ to $100$ (step $5$) \\
        Range of $\lambda$ & $-500$ to $500$ (step $50$) \\
        Value of $\varepsilon$ & $0.05$\\
        \hline
    \end{tabular}
    \label{tab:research_conditions_detail}
\end{table}

In this study, the Fixstars Amplify Annealing Engine (AE), known as an Ising-machine solver, was used to solve the problems~\cite{amplify_engine}.
The problem instances `\texttt{r\_100\_50\_5}', `\texttt{r\_200\_50\_9}', and `\texttt{r\_300\_50\_4}' from the QKP benchmark set were used~\cite{SoutifQKP} (Table~\ref{tab:qkp_instances}).
Each instance consists of the number of items $N=100, 200, 300$ and the interaction matrix density $P_{\mathrm{density}}=50\%$.
Here, $P_{\mathrm{density}}$ denotes the ratio of nonzero elements to the total number of off-diagonal elements in the upper triangular part of Eq.~\eqref{eq:density_matrix}.
The nonzero coefficients $p_{i,j}$ are determined according to a uniform distribution in the range from $1$ to $100$, and the weight of each item is determined according to a uniform distribution in the range from $1$ to $50$.

\begin{table}[t]
\centering
\caption{QKP benchmark instances used in this study.}
\begin{tabular}{l c c c c c}
\hline
Instance & $N$ & $P_{\mathrm{density}}$ [\%] & $W$ & $D$ & $\lvert H_{\mathrm{obj}}^\ast\rvert$\\
\hline
\texttt{r\_100\_50\_5} & 100 & 50 & 983 & 10 & 56464\\
\texttt{r\_200\_50\_9} & 200 & 50 & 1105 & 11 & 104936\\
\texttt{r\_300\_50\_4} & 300 & 50 & 1957 & 11 & 307124\\
\hline
\end{tabular}
\label{tab:qkp_instances}
\end{table}

Because $\mathrm{TT\varepsilon}$ is used as the evaluation metric in this study, problem instances whose optimal solutions are known were adopted.
This makes it possible to strictly evaluate the relative error of the solutions.

The first experiment investigates the $\mu, \lambda$ dependence of $\mathrm{TT\varepsilon}$.
This aims to quantitatively evaluate the degree of influence of $\mu$ and $\lambda$ in the quadratic knapsack problem.
The specific computational conditions are shown in Table~\ref{tab:research_conditions_detail}.
In this study, $p_{R}=0.99$ and the relative error $\varepsilon=0.05$ (within $5\%$ of the optimal solution) are set.
The last column of Table~\ref{tab:qkp_instances} lists the magnitudes of the optimal objective values. Since $H_{\mathrm{obj}}$ is defined in the minimization form in Eq.~\eqref{eq:knapsack_obj}, the corresponding signed optimal value used in the calculations is $H_{\mathrm{obj}}^\ast=-\lvert H_{\mathrm{obj}}^\ast\rvert$.
The execution time in the table corresponds to the annealing time $\tau$ introduced in Subsect.~\ref{subsec:TTepsilon}.
Although $\mu$ is positive in the standard penalty function and augmented Lagrangian function formulations, $\mu=0$ was included in the grid search as a reference limiting case.

The success trials in this study are defined as follows.
The first condition is that the decision variables $x_i$ of the output solution satisfy the constraint $\sum_{i=1}^{N} w_i x_i \leq W$ of the original quadratic knapsack problem.
The second condition is that the relative error based on the objective function $H_{\mathrm{obj}}$, which does not include the constraint term, is at most $\varepsilon$ with respect to the optimal objective value.
A trial that satisfies these two conditions is regarded as a success trial, and $p_{\varepsilon}(\tau)$ introduced in Subsect.~\ref{subsec:TTepsilon} is defined as the empirical probability of such success trials.
Whether the obtained solution satisfies the constraint is judged based on whether the total weight of the items selected by the decision variables $x_i$ is at most the capacity upper limit $W$ ($\sum_{i=1}^{N} w_i x_i \leq W$).
That is, if the combination of decision variables satisfies the constraint, it is regarded as a feasible solution.

Feasibility was evaluated with respect to the original inequality constraint of the QKP, not with respect to the encoded equality condition $H_{\mathrm{const}}=0$. 
This distinction is important because the auxiliary variables are introduced only for QUBO encoding and are not part of the original decision variables.
Therefore, whether the value represented by $y_d$ matches the actual remaining capacity is not included in the success criterion.

The second experiment investigates the temporal evolution of the solver-reported incumbent solution during a single run.
The Fixstars Amplify AE has a function that records the incumbent-solution history at specific intervals within the set execution time, up to the output solution.
This function is used to compare and evaluate the influence of the penalty function and augmented Lagrangian function formulations on the temporal evolution of the incumbent solution.
Because the optimal objective value of each instance used in this experiment is known, the obtained objective function values were normalized by the optimal objective value for evaluation~\cite{parizy2022driving}.
The objective value was normalized as
\begin{equation}
\frac{H_{\mathrm{obj}}(\boldsymbol{x}(t))}{H_{\mathrm{obj}}^\ast},
\label{eq:normalized_obj}
\end{equation}
where both $H_{\mathrm{obj}}(\boldsymbol{x}(t))$ and $H_{\mathrm{obj}}^\ast$ are evaluated in the minimization form of Eq.~\eqref{eq:knapsack_obj}, and $\boldsymbol{x}(t)$ is the incumbent solution at elapsed time $t$.
The parameter search was performed on a finite grid, with steps of 5 in $\mu$ and 50 in $\lambda$. 
Therefore, the minimum $\mathrm{TT\varepsilon}$ reported below should be interpreted as the minimum over the searched grid points, rather than the true minimum in the continuous $\lambda-\mu$ space.

\section{Results}
\label{sec:results}

This section presents the results obtained in this study.
Subsection~\ref{subsec:Time-to-epsilon} presents the hyperparameter dependence of $\mathrm{TT\varepsilon}$.
The following Subsect.~\ref{subsec:solution-update} presents the comparison of the evolution of the incumbent solution under specific hyperparameter settings, based on the findings obtained from the heatmaps.

\subsection{$\mathrm{TT\varepsilon}$ ($\varepsilon=0.05$)}
\label{subsec:Time-to-epsilon}

\begin{figure*}[h]
    \centering
    \begin{minipage}[t]{0.32\linewidth}
        \centering
        \includegraphics[width=\linewidth]{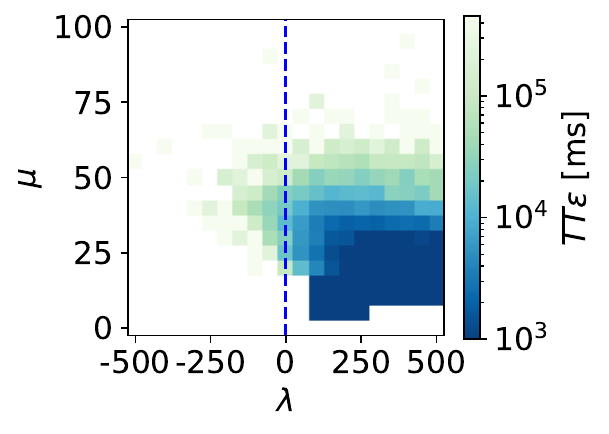}\\
        (a)
    \end{minipage}
    \hfill
    \begin{minipage}[t]{0.32\linewidth}
        \centering
        \includegraphics[width=\linewidth]{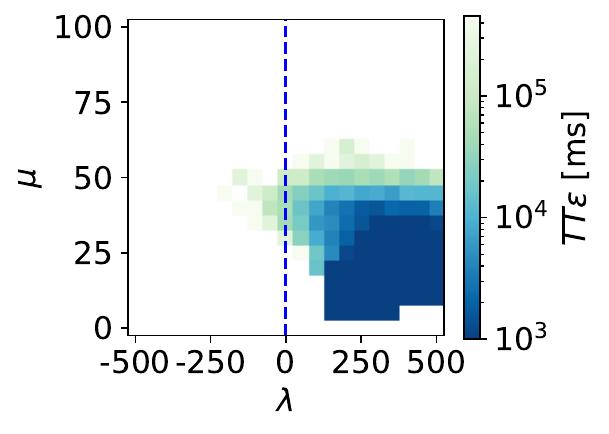}\\
        (b)
    \end{minipage}
    \hfill
    \begin{minipage}[t]{0.32\linewidth}
        \centering
        \includegraphics[width=\linewidth]{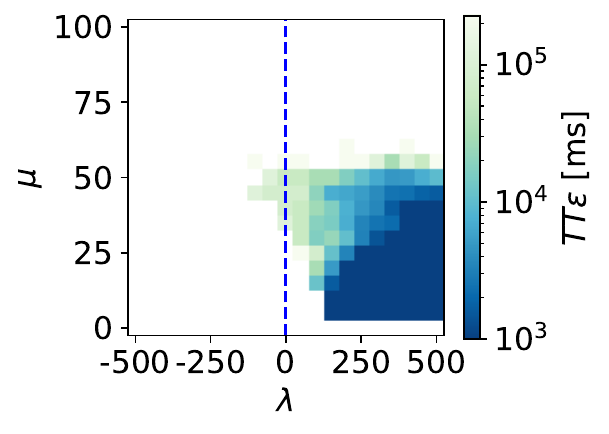}\\
        (c)
    \end{minipage}
    \caption{Distribution of $\mathrm{TT\varepsilon}$ $(\varepsilon=0.05)$ in the $\lambda-\mu$ space for the problem instances treated in this study. Panels (a), (b), and (c) correspond to the instances `\texttt{r\_100\_50\_5}' ($W = 983$), `\texttt{r\_200\_50\_9}' ($W = 1105$), and `\texttt{r\_300\_50\_4}' ($W = 1957$), respectively. The blue line corresponds to the penalty function formulation ($\lambda=0$). The computation was performed under identical conditions $50$ times each, with $\mu$ in steps of $5$ over the range from 0 to 100 and $\lambda$ in steps of $50$ over the range from $-500$ to 500. The white region indicates the region where no solution within a relative error of $\varepsilon$ from the optimal objective value was obtained.}
    \label{fig:TT005_all}
\end{figure*}

This subsection describes the influence of the set hyperparameters on $\mathrm{TT\varepsilon}$, the evaluation metric.
Figures~\ref{fig:TT005_all}~(a), (b), (c) show the heatmaps of $\mathrm{TT\varepsilon}$ in the $\lambda-\mu$ space for each problem instance.
Because the single-run time was fixed at $\tau=1\,\mathrm{s}$, the variation in $\mathrm{TT\varepsilon}$ over the grid reflects the variation in the empirical success probability $p_{\varepsilon}(\tau)$, except at grid points where $p_{\varepsilon}(\tau)=1$, for which $\mathrm{TT\varepsilon}=\tau$, and where $p_{\varepsilon}(\tau)=0$, for which $\mathrm{TT\varepsilon}$ is undefined.
In all instances, a common qualitative tendency was confirmed, in which $\mathrm{TT\varepsilon}$ takes low values in the region where $\lambda > 0$ and $\mu$ is small.
On the other hand, the grid point giving the minimum $\mathrm{TT\varepsilon}$ differed among the three instances, taking $\lambda=100$ for `\texttt{r\_100\_50\_5}' and $\lambda=150$ for `\texttt{r\_200\_50\_9}' and `\texttt{r\_300\_50\_4}'.
A similar qualitative tendency was observed for the three QKP instances tested in this study.
Therefore, the following discussion focuses on the results for the problem instance `\texttt{r\_100\_50\_5}'.

The white grid points indicate parameter settings for which no successful trial was observed among the 50 runs, where a successful trial satisfies both the original QKP constraint and an objective error of at most $\varepsilon$.
For these points, $p_\varepsilon(\tau)=0$ and $\mathrm{TT\varepsilon}$ is undefined.
The region of $\lambda = 0$, indicated by the blue dotted line, corresponds to the conventional penalty function.
Focusing on this region, it can be seen that the range of $\mu$ for which a solution satisfying the relative error $\varepsilon$ can be obtained is limited.
This is because, if $\mu$ is too small, the constraint cannot be satisfied, whereas if $\mu$ is too large, the objective error of the obtained solution increases.
Under the present conditions, the minimum $\mathrm{TT\varepsilon}$ for the penalty function was confirmed at $\mu=40$.

Next, the results for the augmented Lagrangian function formulation are described.
At the grid points $(\mu, \lambda) = (5, 100)$ and $(10, 200)$, $\mathrm{TT\varepsilon}$ was approximately one-tenth of the minimum value obtained for the penalty function formulation.
The heatmap for `\texttt{r\_100\_50\_5}' [Fig.~\ref{fig:TT005_all}~(a)] exhibits a pronounced asymmetry with respect to $\lambda$.
This is because $\lambda > 0$ acts in the direction of satisfying the constraint, whereas $\lambda < 0$ acts in the direction of relaxing the constraint.
Within the positive-$\lambda$ region where successful trials were observed, smaller values of $\mu$ tended to give lower $\mathrm{TT\varepsilon}$.

This observation is consistent with the hypothesis described in Subsubsect.~\ref{subsubsec:augmented_Lagrangian_function}, namely that the introduction of $\lambda$ allows the penalty coefficient $\mu$ required to obtain a feasible solution to be kept relatively small. Conversely, a negative $\lambda$ relaxes the constraint, so that a feasible solution with small objective error is correspondingly hard to obtain. A possible mechanism behind this behavior is discussed in Sect.~\ref{sec:discussion}.

\subsection{Temporal Evolution of the Incumbent Solution}
\label{subsec:solution-update}
For the problem instance `\texttt{r\_100\_50\_5}', based on the evaluation results of $\mathrm{TT\varepsilon}$, the temporal evolution of the incumbent solution within the execution time is compared between the penalty function and augmented Lagrangian function formulations.
The execution time in this study refers not to the total computation time, but to the single-run time $\tau$ allocated to the solver for each annealing run.

For the penalty function formulation, $\mu=40$, for which the result of $\mathrm{TT\varepsilon}$ was small, and its neighboring values $\mu=20, 60$ were selected.
For the augmented Lagrangian function formulation, the hyperparameters $(\mu, \lambda)=(5, 100), (10, 200)$, for which the result of $\mathrm{TT\varepsilon}$ was small, were selected.
The reason for selecting the region with small $\mu$ in the augmented Lagrangian function formulation is to confirm how the property that $\mu$ can be kept small through the introduction of the Lagrange multiplier term influences the temporal evolution of the incumbent solution.
By comparing, for each method, the hyperparameter settings that yielded a small $\mathrm{TT\varepsilon}$ in this way, the influence of the formulation structure itself on the search process is evaluated.

\begin{figure}[hbtp]
    \centering
    \includegraphics[width=\linewidth]{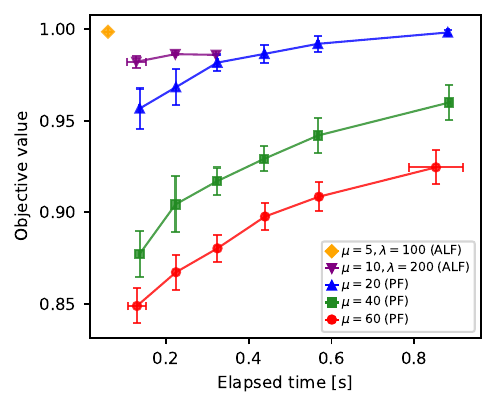}
    \caption{Temporal evolution of the incumbent solution within the execution time of $1\,\mathrm{s}$ in the Fixstars Amplify AE. ALF denotes the augmented Lagrangian function (ALF), and PF denotes the penalty function (PF). The error bars in the time-axis (horizontal) direction arise from the fact that the timing of data acquisition is not strictly identical across trials.}
    \label{fig:solution_update}
\end{figure}

Figure~\ref{fig:solution_update} shows the temporal evolution of the incumbent solution within the execution time.
The horizontal axis represents the elapsed time from the start of execution, and the vertical axis represents the objective function value normalized by the optimal objective value, as defined in Eq.~\eqref{eq:normalized_obj}.
Each curve shows the normalized objective value of the incumbent solution averaged over the feasible solutions obtained across the trials, and the error bars in the vertical direction represent the standard deviation. Only feasible incumbent solutions were included in this evaluation.

In the search process using the penalty function, the incumbent solution is updated stepwise toward better solutions as time elapses, and the solution search ends near the end of the set execution time.
The smaller $\mu$ is, the higher the normalized objective value at both the initial-solution stage and the final solution.
This is because setting $\mu$ small relatively maintains the energy scale of the objective function term relative to the constraint term in the problem Hamiltonian.

When the augmented Lagrangian function formulation is used, under the selected representative hyperparameter settings, the curves for the ALF reached a plateau earlier than those for the PF.
In addition, a major feature is that the variation among the $10$ feasible solutions obtained over multiple trials is small compared with the penalty function. 
Table~\ref{tab:total_trials} shows the number of solver runs required to obtain the $10$ feasible solutions for each parameter setting.

The observations in this subsection are limited to the representative parameter settings and the QKP instance analyzed here.

\begin{table}[t]
  \centering
  \caption{Number of solver runs required to obtain 10 feasible solutions for each parameter setting}
  \label{tab:total_trials}
  \begin{tabular}{ccc}
    \hline
    $\mu$ & $\lambda$ & Total trials \\
    \hline
    5  & 100 & 10 \\
    10 & 200 & 10 \\
    20 & 0   & 66 \\
    40 & 0   & 43 \\
    60 & 0   & 31 \\
    \hline
  \end{tabular}
\end{table}
\section{Discussion}
\label{sec:discussion}

This section discusses the results presented in Sect.~\ref{sec:results}. 
Subsection~\ref{subsec:formulation_change} examines, from the structural change in the formulation, the mechanism by which the augmented Lagrangian function reduces $\mathrm{TT\varepsilon}$, and Subsect.~\ref{subsec:constraint_violation} examines, as an auxiliary analysis, whether the augmented Lagrangian method can reach the favorable parameter region without a prior parameter search.

\subsection{Change in the Formulation}
\label{subsec:formulation_change}
We discuss the reduction of $\mathrm{TT\varepsilon}$ from the perspective of the structural change in the formulation.
For $\mu > 0$, completing the square of Eq.~\eqref{eq:knapsack_ALM} gives
\begin{equation}
    H_{\mathrm{AL}} = H_{\mathrm{obj}} + \frac{\mu}{2}\left(H_{\mathrm{const}} - \frac{\lambda}{\mu}\right)^2 - \frac{\lambda^2}{2\mu}.
    \label{eq:AL_completing_the_square}
\end{equation}
The last term, $-{\lambda^2}/{2\mu}$, is independent of the binary variables and does not affect the minimizer of the Hamiltonian.
From Eq.~\eqref{eq:AL_completing_the_square}, the addition of the Lagrange multiplier term $\lambda$ has the effect of shifting the energy minimum of the penalty term to $H_{\mathrm{const}} = {\lambda}/{\mu}$.

This shift is mathematically equivalent to directly shifting the constraint term of the penalty function, that is, to introducing a shift amount $s$ as in $\frac{\mu}{2}(H_{\mathrm{const}} - s)^2$.
However, the two formulations differ essentially in how the shift amount is given.
The shift amount $s$ must be fixed in advance as a new parameter, and its appropriate value is unknown and problem-dependent.
This involves a difficulty similar to that of setting the penalty coefficient $\mu$.
By contrast, in the augmented Lagrangian function, the shift amount is given as ${\lambda}/{\mu}$ in terms of the Lagrange multiplier $\lambda$.
The Lagrange multiplier $\lambda$ is a quantity that can be updated based on the amount of constraint violation, and the augmented Lagrangian method described later can determine the shift amount adaptively.
That is, the advantage of the augmented Lagrangian function lies not in shifting the energy minimum itself, but in providing a framework that determines the shift amount based on the constraint violation.

Using Eq.~\eqref{eq:knapsack_const}, the condition $H_{\mathrm{const}} = {\lambda}/{\mu}$ can be rewritten as
\begin{equation}
    \sum_{i=1}^{N} w_i x_i + \left(2^D - 1 - \sum_{d=0}^{D-1} 2^d y_d\right) = W - \frac{\lambda}{\mu}.
    \label{eq:effective_capacity_shift}
\end{equation}
Thus, for $\lambda > 0$, the encoded equality condition is shifted toward a smaller effective capacity.
This shift biases the search toward item selections with smaller total weight and can increase the probability of satisfying the original QKP inequality.

The influence of shifting the energy minimum to $H_{\mathrm{const}} = {\lambda}/{\mu} > 0$ is also described from the perspective of the auxiliary variables $y_d$.
From Eq.~\eqref{eq:knapsack_const}, satisfying $H_{\mathrm{const}} = {\lambda}/{\mu}$ requires the sum of the auxiliary variables $\sum_{d=0}^{D-1} 2^d y_d$ to take a value larger than its original value.
For a feasible item selection with actual remaining capacity $r = W - \sum_{i=1}^{N} w_i x_i$, the auxiliary variables encode $\sum_{d=0}^{D-1} 2^d y_d = 2^D - 1 - r$ when $H_{\mathrm{const}} = 0$.
After the shift, the required encoded value becomes $2^D - 1 - r + ({\lambda}/{\mu})$.
When ${\lambda}/{\mu} > r$, this value exceeds the maximum representable value $2^D - 1$, and the auxiliary variables are driven to the saturated assignment $y_d = 1$ for all $d$.
In this saturated state, the encoded remaining capacity $2^D - 1 - \sum_{d=0}^{D-1} 2^d y_d$ becomes zero and no longer matches the actual remaining capacity $W - \sum_{i=1}^{N} w_i x_i$.
That is, the auxiliary variables need not represent the actual remaining capacity even when the item-selection variables satisfy the original inequality constraint.

\begin{figure}[t]
    \centering
    \includegraphics[width=0.8\linewidth]{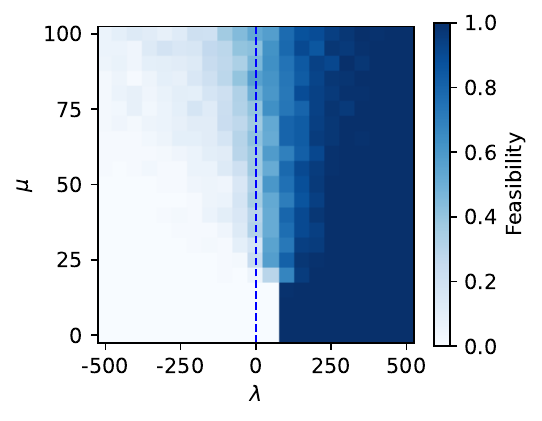}
    \caption{Distribution of the feasibility in the $\lambda-\mu$ space for the problem instance `$\mathrm{r\_100\_50\_5}$' treated in this study. The feasibility is the empirical probability that the returned item-selection variables satisfy the original QKP constraint. The weight is $W = 983$. When the feasibility becomes $1.0$ at $\mu=0$ due to the addition of the Lagrange multiplier term, the obtained solution is the trivial solution in which no item is selected.}
    \label{fig:feasible_solution_heatmap}
\end{figure}

This discussion is also consistent with Fig.~\ref{fig:feasible_solution_heatmap}, which shows the feasibility in the $\lambda-\mu$ space.
Here, the feasibility is defined as the empirical probability that the returned item-selection variables satisfy the original QKP constraint, irrespective of the objective error.
It therefore differs from the success probability $p_{\varepsilon}(\tau)$, which additionally requires the objective error to be at most $\varepsilon$.
In the region of small $\mu$ where the penalty function never yields a feasible solution, the introduction of the Lagrange multiplier term raises the feasibility to $1.0$.
The increase in feasibility alone does not imply an improvement in solution quality.
In particular, at $\mu = 0$, the feasibility can become $1.0$ because the trivial solution with no selected item is returned.
The reduction of $\mathrm{TT\varepsilon}$ should therefore be interpreted as the combined effect of increased feasibility and the maintained contribution of the objective term.

On the other hand, not only constraint satisfaction but also the objective function value of the solution contributes to the improvement of the solution performance.
As a rough diagnostic of the coefficient scale before squaring the constraint function, we compare the norms of the coefficients in $H_{\mathrm{const}}$ and $H_{\mathrm{obj}}$, each defined as the square root of the sum of the squares of all coefficients.
The coefficients of $H_{\mathrm{obj}}$ are $p_{i,j}$, and the coefficients of $H_{\mathrm{const}}$ are the item weights $w_i$ and the auxiliary-variable weights $2^d$.
For the problem instance `\texttt{r\_100\_50\_5}' treated in this study, the norm ratio of the two is given by the following equation.
\begin{equation}
    \frac{\lVert H_{\mathrm{const}} \rVert}{\lVert H_{\mathrm{obj}} \rVert}
    = \frac{\sqrt{\displaystyle\sum_{i=1}^{N} w_i^2 + \sum_{d=0}^{D-1} \left(2^d\right)^2}}
           {\sqrt{\displaystyle\sum_{1 \le i \le j \le N} p_{i,j}^2}}
    \approx 0.22.
    \label{eq:norm_ratio}
\end{equation}
This rough comparison indicates that the two coefficient scales are of similar order before squaring.

However, in the problem Hamiltonian, the constraint term is given not as $H_{\mathrm{const}}$ itself but as $\frac{\mu}{2}(H_{\mathrm{const}})^2$, as shown in Eq.~\eqref{eq:knapsack_PM}.
For this reason, the effective energy scale of the constraint term relative to the objective function term depends on both the magnitude of $H_{\mathrm{const}}$ and the value of $\mu$, and increases with both accordingly.
Therefore, keeping $\mu$ small relatively maintains the energy scale of the objective function term relative to the constraint term.
This makes feasible solutions with small objective error easier to obtain.
This is consistent with the result in Fig.~\ref{fig:TT005_all}~(a), where $\mathrm{TT\varepsilon}$ is minimized in the region with $\lambda > 0$ and small $\mu$.
From the above, the structural change in the formulation by the augmented Lagrangian function is considered to contribute to the improvement of the solution performance of the Ising machine.

\subsection{Constraint Violation}
\label{subsec:constraint_violation}
In Subsect. \ref{subsec:Time-to-epsilon}, the hyperparameters $\mu, \lambda$ were set to various values, and their space was exhaustively explored.
This clarified the region in which $\mathrm{TT\varepsilon}$ is minimized.
However, in practical solving, a low-$\mathrm{TT\varepsilon}$ region cannot be known in advance.
The augmented Lagrangian method is an algorithm that adaptively updates $\mu, \lambda$ based on the amount of constraint violation.
This subsection examines, as an auxiliary analysis, whether parameter updates based on the augmented Lagrangian method rule can move the parameters toward the low-$\mathrm{TT\varepsilon}$ region identified in Subsect.~\ref{subsec:Time-to-epsilon}.
In the penalty method described in Subsubsect. \ref{subsubsec:penalty_method}, as shown in Eq.~\eqref{eq:mu_update}, $\mu$ is increased uniformly by a fixed growth factor $\alpha$.
By contrast, the augmented Lagrangian method dynamically updates the Lagrange multiplier $\lambda$ based on Eq.~\eqref{eq:lambda_update}, according to the amount of constraint violation of the obtained solution.
In this subsection, the constraint violation is measured by $-H_{\mathrm{const}}$, where $H_{\mathrm{const}}$ is defined in Eq.~\eqref{eq:knapsack_const}. This quantity is zero when the constraint is satisfied and positive when it is violated.

The hyperparameter dependence of the constraint violation in the $\lambda-\mu$ space is investigated.
In this investigation, the hyperparameters were set on a finer grid than that used for Fig.~\ref{fig:TT005_all}~(a), and the computation was performed again.
Here, $\mu$ was set in steps of $1$ over the range from 1 to 40, and $\lambda$ in steps of $5$ over the range from 0 to 150.
In addition, for the hyperparameter update trajectories of the augmented Lagrangian method, the initial value $\mu^{(0)}$ and the growth factor $\alpha$ were set under multiple conditions, and the computation was performed.
The initial value $\lambda^{(0)}$ was commonly set to $\lambda^{(0)}=0$.
At each step, the hyperparameters were retained until at least one feasible solution, in the sense of the original QKP constraint, was obtained.
The hyperparameters were updated according to the update rule only when no feasible solution was obtained.

\begin{figure*}[t]
    \centering
    \begin{minipage}[t]{0.48\linewidth}
        \centering
        \includegraphics[width=\linewidth]{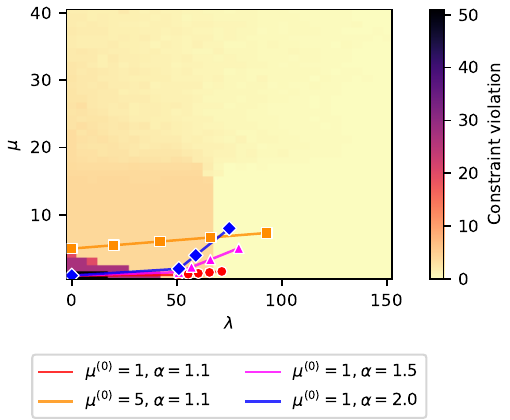}\\
        (a)
    \end{minipage}
    \hfill
    \begin{minipage}[t]{0.48\linewidth}
        \centering
        \includegraphics[width=\linewidth]{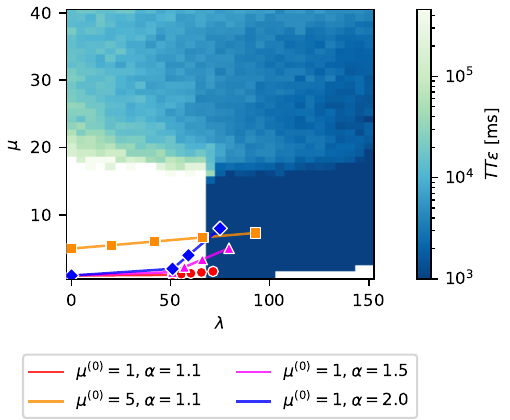}\\
        (b)
    \end{minipage}
    \caption{Distributions in the $\lambda-\mu$ space for the problem instance `\texttt{r\_100\_50\_5}' ($W = 983$). Panel (a) shows the constraint violation, and panel (b) shows $\mathrm{TT\varepsilon}$ $(\varepsilon=0.05)$. The computation was performed under identical conditions $50$ times each, with $\mu$ in steps of $1$ over the range from $1$ to $40$ and $\lambda$ in steps of $5$ over the range from $0$ to $150$. In panel (a), the constraint violation at each point is the average value of the constraint violation obtained over $50$ trials under the same hyperparameter setting. The trajectories in both panels show the parameter transitions when the hyperparameters are updated according to the augmented Lagrangian method, and the starting point of each trajectory corresponds to the initial value $\mu^{(0)}$ and $\lambda^{(0)}=0$.}
    \label{fig:100_50_5_const_violation_TT}
\end{figure*}

Figure~\ref{fig:100_50_5_const_violation_TT}~(a) shows the constraint violation for the above hyperparameter settings, together with the parameter update trajectories of the augmented Lagrangian method starting from the initial values $\mu^{(0)}$ and growth factors $\alpha$ under multiple conditions.
The constraint violation here is the average value of the constraint violation obtained over $50$ trials for each hyperparameter setting.
From Fig.~\ref{fig:100_50_5_const_violation_TT}~(a), it can be seen that the constraint violation is large in the region where $\mu$ and $\lambda$ are close to $0$.
Therefore, when $\mu^{(0)}=1$ is set, $\lambda$ is updated by a large amount in the first few steps.
Focusing on the trajectories, it can also be seen that, even when starting from multiple initial values $\mu^{(0)}$ and growth factors $\alpha$, the hyperparameters are updated in the direction in which the constraint violation decreases.

The heatmap of $\mathrm{TT\varepsilon}$ under the same conditions is shown in Fig.~\ref{fig:100_50_5_const_violation_TT}~(b).
From this figure, it can be seen that, in the augmented Lagrangian method, the hyperparameters are updated toward the region in which $\mathrm{TT\varepsilon}$ is minimized.
In particular, for the setting near the growth factor $\alpha=1.1$, which is considered appropriate for the augmented Lagrangian method~\cite{djidjev2023quantum}, the region in which $\mathrm{TT\varepsilon}$ is minimized was reached while keeping the value of $\mu$ small.
This supports the claim of a previous study~\cite{tanahashi2021augmented}.
In this study, the augmented Lagrangian function was evaluated as a formulation method, and its advantage was shown.
These results suggest that the advantage identified in Sect.~\ref{sec:results} can potentially be exploited through the augmented Lagrangian method, which updates the parameters based on the amount of constraint violation.
These results suggest that, for the tested instance and update settings, ALM-type updates can move the parameters toward the low-$\mathrm{TT\varepsilon}$ region without using the full grid-search information.
A systematic comparison of the total cost of such iterative updates, which require multiple solver calls, with that of a direct parameter search remains future work.

The above discussion explains the reduction of $\mathrm{TT\varepsilon}$ mainly through the shift of the encoded constraint condition and the resulting increase in feasibility at small $\mu$.
However, it does not fully clarify why the incumbent objective value reaches its plateau earlier for the augmented Lagrangian function, as observed in Subsect.~\ref{subsec:solution-update}.
Clarifying this dynamical aspect requires a more detailed analysis of the solver-reported state trajectories.
In addition, the present discussion is based on QKP instances from a single benchmark family, and the generality to other constrained combinatorial optimization problems remains to be examined.

\section{Conclusion}
\label{sec:conclusion}
In this study, the augmented Lagrangian function was evaluated as a formulation for the Ising machine, using $\mathrm{TT\varepsilon}$ as the evaluation metric.
For the QKP, which is an inequality-constrained combinatorial optimization problem, the dependence of $\mathrm{TT\varepsilon}$ on the hyperparameters $\mu$ and $\lambda$ was investigated using the benchmark instances listed in Table~\ref{tab:qkp_instances}.
The investigation showed that the augmented Lagrangian function formulation yielded a low $\mathrm{TT\varepsilon}$ in a specific parameter region, about one order of magnitude lower than the minimum value of the conventional penalty function formulation.
Furthermore, the temporal evolution of the incumbent solution within the execution time was investigated.
This revealed that, for the representative parameter settings selected in this study, the incumbent objective value among feasible runs approached the optimal objective value earlier in the augmented Lagrangian function formulation than in the penalty function formulation.

The square-completion analysis suggests that positive $\lambda$ shifts the encoded constraint condition toward a smaller effective capacity. This shift provides a possible explanation for the observed reduction of $\mathrm{TT\varepsilon}$, because it increases the probability of sampling feasible item selections while keeping $\mu$ small.
The present conclusions are limited to the QKP benchmark instances and solver settings examined in this study.
On the other hand, the detailed mechanism that brings about the early asymptotic approach to the value of the optimal objective value remains unclarified.
A future task is to investigate the influence of the introduction of the Lagrange multiplier term on the transition behavior between states in the solution search space, and thereby to clarify the mechanism of this early asymptotic approach.

\section*{Acknowledgments}
This work was partially supported by the Japan Society for the Promotion of Science (JSPS) KAKENHI (Grant Number JP23H05447), the Council for Science, Technology, and Innovation (CSTI) through the Cross-ministerial Strategic Innovation Promotion Program (SIP), ``Promoting the application of advanced quantum technology platforms to social issues'' (Funding agency: QST), Japan Science and Technology Agency (JST) (Grant Number JPMJPF2221). S. Tanaka wishes to express gratitude to the World Premier International Research Center Initiative (WPI), MEXT, Japan, for supporting the Human Biology-Microbiome-Quantum Research Center (Bio2Q).

\bibliographystyle{jpsj}

\bibliography{reference}

\end{document}